\documentclass[10pt, journal]{IEEEtran}
\IEEEoverridecommandlockouts

\usepackage{subfigure}

\usepackage[utf8]{inputenc}
\usepackage{amsmath,amssymb}

\usepackage{graphicx}
\usepackage{subfigure}
\usepackage{color}
\usepackage{array}
\usepackage{url}

\usepackage{algorithm}
\usepackage{algorithmic}

\begin{document}

\title{Parallel and Distributed Simulation\\ from Many Cores to the Public Cloud\\(Extended Version)\footnotemark}

\author{\IEEEauthorblockN{Gabriele D'Angelo}
\IEEEauthorblockA{\\Department of Computer Science\\ University of Bologna\\
Bologna, Italy\\
g.dangelo@unibo.it,\\
\url{http://www.cs.unibo.it/gdangelo}}
}

%\IEEEpubid{\makebox[\columnwidth]{\hfill
%978-1-4244-9538-2/11/\$26.00~\copyright~2011 IEEE}
%\hspace{\columnsep}\makebox[\columnwidth]{}}

\maketitle

\footnotetext{The publisher version of this paper is available at \url{http://dx.doi.org/10.1109/HPCSim.2011.5999802}.
\textbf{{\color{blue}A journal version of this paper has been published in \url{http://dx.doi.org/10.1016/j.simpat.2014.06.007}}}.
\textbf{{\color{red}Please cite this conference version as: Gabriele D'Angelo. Parallel and Distributed Simulation from Many Cores to the Public Cloud. Proceedings of the 2011 International Conference on High Performance Computing and Simulation (HPCS 2011), Istanbul (Turkey), IEEE, July 2011. ISBN 978-1-61284-382-7.}}}

\begin{abstract}
In this tutorial paper, we will firstly review some basic simulation concepts and then introduce the parallel and distributed simulation techniques in view of some new challenges of today and tomorrow. More in particular, in the last years there has been a wide diffusion of many cores architectures and we can expect this trend to continue. On the other hand, the success of cloud computing is strongly promoting the ``everything as a service'' paradigm. Is parallel and distributed simulation ready for these new challenges? The current approaches present many limitations in terms of usability and adaptivity: there is a strong need for new evaluation metrics and for revising the currently implemented mechanisms. In the last part of the paper, we propose a new approach based on multi-agent systems for the simulation of complex systems. It is possible to implement advanced techniques such as the migration of simulated entities in order to build mechanisms that are both adaptive and very easy to use. Adaptive 
mechanisms are able to significantly reduce the communication cost in the parallel/distributed architectures, to implement load-balance techniques and to cope with execution environments that are both variable and dynamic. Finally, such mechanisms will be used to build simulations on top of unreliable cloud services.
\end{abstract}


\section{Introduction}

% Introduzione alla simulazione
A computer simulation is a computation that models the behavior of some real or imagined system over time~\cite{FUJ00}. In practice, it consists in a set of techniques that are fundamental for the performance evaluation of existing systems, for the study of new solutions and for the creation of virtual words (e.g.~online games, digital virtual environments). 

% Motivazioni al suo utilizzo
There are many reasons behind the use of simulation techniques, some of which are quite convincing. For example, the system that needs to be evaluated can not be built (e.g.~for cost reasons), testing on an existing system can be very dangerous (and some stress testing is actually impossible to perform), and often it is necessary to explore many different solutions in order to choose the best one. The demand for more and more complex systems has led to a wide diffusion of the simulation techniques, a large amount of research on this field and the availability of many different software tools.\\

% A few words about discrete event simulation
In the years, many simulation paradigms have been proposed, each one presenting some benefits and drawbacks. Among them, the Discrete Event Simulation (DES)~\cite{Law:1999:SMA:554952} is powerful in terms of expressiveness and easy to understand for the developer of simulation models. In a DES, the evolution of a modeled system is represented as a chronological sequence of events. Each event represents a change in the system state and occurs at an instant in time. Hence, the evolution of a system from the bootstrap to the end is obtained through the creation, delivery and computation of events. For example, in the simulation of mobile wireless devices, some events will be the transmission of data packets and the motion of devices. In its simplest form, a DES is implemented using a set of state variables (i.e.~to describe the modeled system), an event list (i.e.~the pending events that have been generated by the processing of simulated events and that will be computed in future), and a global clock (i.e.~the 
current simulation time)~\cite{Law:1999:SMA:554952}.\\

When all such tasks are accomplished by a single execution unit (e.g.~a CPU and some random access memory), that is a sequential (i.e.~monolithic) simulator. That means that such single execution unit is responsible for the modeling of the whole system and the management of its evolution, and to do this it processes all the generated events, in sequential order. The main advantage of this approach is its simplicity, but it also introduces some severe limitations: the memory resources of a single execution unit can be insufficient for the task of modeling complex systems. Furthermore, the amount of time needed to complete the simulation runs can be excessive~\cite{1668384}.\\

An alternative approach, called Parallel Discrete Event Simulation (PDES)~\cite{Fujimoto:1989:PDE:76738.76741}, relies on multiple interconnected execution units (e.g.~CPUs or hosts). In this case, each execution unit manages only a part of the simulated model. Thus, it is possible to represent very large and complex models using aggregated resources from many execution units. Differently from DES, in PDES each execution unit has to manage its local event list and locally generated events may have to be delivered to remote execution units. Furthermore, their processing has to be synchronized with the rest of the simulator. As said before, the benefit of using computation and memory resources aggregation is that it permits the simulation of very large and complex systems and, in many cases, the parallel execution of concurrent events~\cite{Lamport1978} can lead to a significant speedup of the simulation execution.\newline

% Paper organization
This paper is organized as follows. In the next section we will provide some background notions about Parallel and Distributed Simulation (PADS). Section \ref{sec:challenges} is about the many cores architecture and the public cloud both of which are new challenges for PADS. The functionality and limitations of current PADS approaches are discussed more in detail in Section \ref{sec:limitations}. In Section \ref{sec:artisgaia} we will describe and discuss our proposal aimed to obtain more adaptable PADS. Finally, Section \ref{sec:conc} will provide some concluding remarks.

%\IEEEpubidadjcol

\section{Parallel and Distributed Simulation}
\label{sec:pads}

% Motivations
Between all, one of the simplest and more general definitions of Parallel and Distributed Simulation (PADS) is: ``any simulation in which more than one processor is employed''~\cite{perumalla2007}. There are many reasons for relying on PADS: to obtain the results faster, to simulate larger scenarios, to integrate simulators that are geographically distributed, to integrate many commercial off-the-shelf simulators and to compose different simulation models in a single simulator~\cite{FUJ00}.\\

% Description of the PADS approach
Under the technical viewpoint, the main difference between sequential simulation and PADS is the lack of a global state that is the representation of the simulated system in a synthetic model. A PADS is obtained through the interconnection of a set of model components, usually called Logical Processes (LPs). Therefore, each LP is responsible to manage the evolution of a part of the system and interacts with the other LPs for all the synchronization and data distribution issues~\cite{FUJ00}. In practice, each LP is usually executed by a processor (or a core in modern multi-core architectures). The type of network that interconnects the processors is of main importance, given that it will strongly affect the simulator characteristics and performance. 
%	parallel or distributed?
The difference between parallel and distributed simulation is quite an elusive one. Usually, the term parallel simulation is used if the processors have access to some shared memory or in presence of a tightly coupled interconnection network. Conversely, we talk about distributed simulation in case of loosely coupled architectures (i.e.~distributed memory)~\cite{perumalla2007}. Obviously, real world execution platforms are very often a mix of the two, as it happens in LAN-based clusters of multi-CPU (and multi-core) hosts.\\
%	consequences of PADS
The lack of a global state and the presence of a network that interconnects the different parts of the simulator has some important consequences:
\begin{itemize}
%		partitioning
	\item the model that represents the simulated system has to be partitioned in components (the LPs)~\cite{bagrodia98}. In some cases, this \textbf{partitioning} is guided by the structure and the semantic of the simulated system (e.g.~if it is composed of some parts, each one having its behavior and structure but interacting with the others). In other cases, the partitioning task is much more complex (e.g.~the system is monolithic and hard to split in parts). In all cases, when partitioning, many different aspects have to be considered. For example, both the minimization of the amount of network communication in the simulator and the load balancing of the execution architecture have a very deep impact on the simulator performance;
%		synchronization
	\item the results of a parallel/distributed simulation are correct only if the outcome is identical to the one that we would have obtained from a sequential one. That's impossible if the PADS does not implement some kind of \textbf{synchronization} among the different parts that compose the simulator. Specific algorithms are needed for the synchronization of the LPs involved in the execution process;
%		data distribution management
	\item each component of the simulator will produce state updates that are possibly relevant for other components. The distribution of such updates in the execution architecture is called \textbf{data distribution}, and for overhead reasons it can not be implemented using broadcasts. The correct approach is to match the data production and consuming based on interest criteria: only the necessary data has to be delivered to the interested components~\cite{Jun:2002:ESM:564062.564074}.
\end{itemize}

\subsection{Synchronization}
% PDES -> PADS -> more details about synchronization algorithms
Implementing a PDES in a PADS architecture requires that all generated events have to be timestamped and delivered following a message-based approach. 
Two events are said to be in causal order if one of them can have some consequences on the other~\cite{Lamport1978}. This constraint is quite easy to satisfy in a sequential simulation: all the events have to be considered in non-decreasing order with respect to their timestamp. In a parallel or distributed architecture the components can proceed at different speed and the network can introduce unpredictable delay and loss in the messages delivery. To guarantee that the PADS does not violate the causality constraint, all the LPs involved in the simulation execution have to be coordinated using some sort of synchronization algorithm.\\

In the last decades, many different approaches and variants have been investigated but, with some simplification, three main methods are used: 
\begin{itemize}
	\item \emph{time-stepped}: the simulated time is divided in fixed-size timesteps and each LP can proceed to the next timestep only when all other LPs have completed the elaboration of the current timestep~\cite{1261535}. Under the implementation point of view, this approach is quite simple but the division in timesteps can be challenging for some simulation models;
	\item \emph{conservative}: the goal of this approach is to prevent causality errors. This means that, before processing an event with timestamp $t$, the LP has to decide if this event is ``safe'' or not. It can be considered ``safe'' if, in the future, there will be no events with timestamp less than $t$. If this rule is followed by all LPs, then the PADS will obtain results that are correct under the synchronization viewpoint. In practical terms, this has been implemented in many different ways. For example, the Chandy-Misra-Bryant algorithm~\cite{misra86} introduces in the simulation some events without any semantic content that are needed to verify if events are ``safe'' and to avoid deadlocks;
	\item \emph{optimistic}: in this case the LPs are free to violate the causality constraint and, for example, to process the events in receiving order. There is no \emph{a priori} attempt to predict the arrival of a new event with lower timestamp that will cause a causality violation. If that will happen, then the LP will have to roll-back to a previous internal state that is considered correct and to propagate the roll-back to the other affected LPs~\cite{timewarp,quaglia2003}.
\end{itemize}

% final sentence about the many different approaches and the great amount of research work on the tuning of them to the different simulated models,
% execution environments, scenarios
In the years, all such approaches have been deeply investigated and many variants have been proposed. More in detail, we have learned that the performance of the synchronization algorithms heavily depends on many factors such as: the simulation model, the execution environment and the specific scenario. Forecasting the performance of a PADS is very hard, given that it depends on so many factors, some of which are static and known in advance while many others are unknown or depend on the runtime conditions.

\subsection{Software tools}
Many tools have been developed to ease the implementation of PADS. Some of them are compliant with the IEEE 1516 - High Level Architecture (HLA) standard~\cite{ieee1516}. A few examples are: RTI NG Pro~\cite{rti-ng-pro}, Georgia Tech FDK~\cite{fdk}, MAK RTI~\cite{mak}, Pitch RTI~\cite{pitch}, CERTI Free HLA~\cite{certi}, OpenSkies Cybernet~\cite{openskies}, Chronos~\cite{chronos} and the Portico Project~\cite{portico}. Many others have more focus on other aspects such as performance, extensibility or development of new techniques. It is not possible to list all of them but some very interesting tools are: $\mu$sik~\cite{Perumalla:2005:MPS:1069810.1070161}, SPEEDES~\cite{Steinman:2003:SPF:824475.825880} and PRIME~\cite{prime}.

\section{New Challenges of Today and Tomorrow}
\label{sec:challenges}

The technological evolution in computing is fast, sometimes confusing, but it possible to identify some characteristics and trends. New features are very often introduced in hardware, but software is slow in supporting them. In software, many techniques and mechanisms are strongly affected by hardware characteristics such as the internals of CPUs, random access memory and networks. For example, for many years, 32 bits processors have limited the amount of memory that can be used by sequential simulators. Nowadays, 64 bits CPUs and operating systems are quite common and, at least theoretically, this limit is much less severe.\\

We find that there are some trends that will heavily shape PADS in the next years and that can not be ignored. The first one concerns microprocessors: the so called ``MHz race'' (e.g.~the very fast increase of clock speed) has slowed down and multi-core processors are available on the market at very affordable prices. This means that in the same integrated circuit die there are two or more independent real processors and that, very often, all such cores can communicate quite efficiently. On the other hand, only few users of simulation tools can access High Performance Computing (HPC) facilities. For cost reasons, many of them are willing to use only Commercial Off-The-Shelf (COTS) hardware and they would frequently like to build execution platforms with hardware that is already used for other tasks (e.g.~desktop PCs or underloaded servers). The next logical step in this direction is the outsourcing of the computing tasks, such as the execution of simulations. One main goal of cloud computing is to offer 
``pay-as-you-go'' virtual computing environments in which you pay only for capacity that you actually use~\cite{amazon}. More precisely, cloud computing is a model for enabling ubiquitous, convenient, on-demand network access to a shared pool of configurable computing resources that can be rapidly provisioned and released with minimal management effort or service provider interaction~\cite{nist-cloud}.\\

%frase di chiusura e poi approfondimento
Many of the available simulators are unable to cope with such changes in the execution environments: at best they will not exploit all the available resources, but it happens more and more often that users will be encouraged to oversimplify the simulation models. A very risky move. In the following of this section, we will discuss more specifically some of these changes.

	\subsection{The many cores architectures}
% finora ho accennato solo dei multi core e non ho detto nulla dei many core
Nowadays, entry-level CPUs provide 2 or 4 cores but processors with up to 16 cores are already available on the market. The next-generation processors will further increase the available cores and CPUs with 100 cores have been already announced to due out in fourth quarter of 2011~\cite{tilera100}. As usual, the many cores architectures will firstly arrive on the server market, and only in the following years they will be used in desktop PCs. Under the simulation viewpoint, this change in the execution architecture will not be transparent to users. Sequential simulators are, for the most part, unable to exploit more than one core. This means that PADS techniques will be necessary even to run simulations on a desktop PC. Furthermore, most of the multi-processors that have been used until now are based on Symmetric MultiProcessing (SMP), in which two or more identical processors are connected to a single shared main memory. Many cores architectures are much less homogeneous: some cores can be dedicated to 
specific tasks and the access to the main memory (and the caches) can be very asymmetric. All such aspects will become very important given that they strongly affect the simulators performance. Following the PADS approach, the availability of a larger amount of cores implies the partitioning of the simulation model in more and more LPs.\\

As said before, the already complex task of model partitioning becomes even harder as the number of partitions increase. Here is what happens: the simulation modeler, who is in charge of this task, needs to know very well (i.e.~to predict) the behavior of the simulated model. This is essential to obtain an adequate partitioning, so that the communications among the partitioned simulation model are minimized and each computation core is not overloaded (\textbf{load balancing}). Both communication and load balancing can easily become a bottleneck. The partitioning task is complicated by the need to make all choices \emph{a priori} (i.e.~before the simulation starts) and by the constraint that all allocations are static (i.e.~they can not be dynamically changed at runtime). It is clear that most users of simulation tools are unable to complete such a complex task in the correct way. However, their goal is to obtain some results about the analyzed systems, not to become an expert of computing architectures or 
PADS techniques.

	\subsection{Simulation as a service: simulation in the public cloud}
	\label{subsec:service}
% indicare che la simulazione può essere vista come un servizio esternalizzabile
% non ci sarà più bisogno di avere alcuna infrastruttura di calcolo interna, risparmio molto notevole
Some research work on the usage of cloud computing technologies for the implementation of PADS has already been done~\cite{fuj2010-cloud1,fuj2010-cloud2,5564701}. The usage of such technologies in next generation computing infrastructures (i.e.~private cloud) is promising, but many simulation users are still more interested in exploiting the existing public cloud infrastructures for the execution of simulation runs. An execution environment where there is no need for any investment in hardware would be much appreciated by a lot of small and medium-size firms that need to perform simulations but have a very limited budget. Following the ``everything as a service'' paradigm, all the computation resources that are necessary for the PADS execution can be rented from the many providers of cloud services available on the market (e.g.~Amazon, Google, Microsoft and so on). The option to pay only for the used resources is very attractive, but even more charming is the ability to increase and decrease dynamically the 
rented resources.\\

Obviously, the approach based on the public clouds can also have many drawbacks: all the \textbf{partitioning problems} described above are still unsolved and furthermore a public cloud environment, due to its nature, is much more unpredictable (in terms of performance) than other, more classic, execution environments (e.g.~multi-processors or clusters). For example, the many virtual instances composing the execution environment can be located in different data centers (and supplied by one or more providers). As part of the Internet, each interconnecting network will be subjected to performance variations. Furthermore, the performance of each virtual instance will be variable and influenced by many factors such as the usage ratio of the hardware machine providing the virtualization service.\\

% in presenza di fault tolerance PADS... utilizzo di servizi cloud non affidabili ma estremamente economici
That brings to an extreme but interesting evolution of this approach. The price of cloud computing services can vary very much depending on factors such as guaranteed performance and reliability. Almost all providers of cloud computing solutions spend a lot in terms of resources with the aim of providing reliable and fault-tolerant infrastructures. For example, offering data centers that are located in many parts of the world and with redundant connectivity. The assumption behind this commercial offer is that all users require the best levels of service for their applications. Focusing on the simulation users, it is worth noting that, in some cases, it is not strictly necessary to obtain the results in real-time (or faster than possible), as users can wait for some extra time before obtaining them. To complete the simulation runs can require many hours, and a little delay is almost negligible. Therefore, an alternative approach would be to add some amount of fault-tolerance in the PADS mechanisms and only 
rent very inexpensive (and low reliability) cloud services. In some cases, this approach could lead to a very significant reduction of the lease cost, at the price of some increment in the time required to complete the simulation runs. Obviously, the middleware used to build the distributed simulation has to implement a replication mechanism, that is necessary for dealing with the faulty parts of the execution architecture. It is clear that there will be no gain if the overhead introduced by the replication mechanism is too high. In Section~\ref{sec:artisgaia} we will see that fault-tolerance and load balancing can be considered in an integrated way and that a migration-based middleware can be used to address both problems.

\section{Functionality and Limitations of current PADS approaches}
\label{sec:limitations}

%
% Modificare il testo quanto basta per renderlo più scorrevole e più omogeneo, alcuni concetti potrebbero essere eventualmente un po' espansi ed eliminare alcune ripetezioni
%
Given the increasing complexity of the studied systems, we would expect a broad application of PADS techniques: this is not the case. It happens in fact that many users are unwilling to dismiss the ``old'' (sequential) tools and switch to more modern ones, despite the very strong demand for scalability and execution speed. What is missing? There is clearly a problem that should be properly defined.\\

Two of the main goals of the last decades research work on PADS were: i) \emph{make it fast}; ii) \emph{make it easy to use}~\cite{fujtut2000}. Today, we can say that PADS, in the right conditions, can be very fast~\cite{Perumalla:2007:STW:1242531.1242543}. Above all, the research work on synchronization algorithms and data distribution management has allowed to increase significantly the speed-up of simulation runs. This is true if the simulation model is properly partitioned among the execution architecture, the appropriate synchronization algorithm is used (each one of them has its characteristics and limitations), and the execution architecture is fast, reliable and in most cases homogeneous (in terms of performance of each node). In other words, the execution speed of PADS is limited by its slowest component and therefore a good control on the whole simulator and its execution architecture is necessary~\cite{gda-ijspm-2009}.
In terms of usability, there is not much more to be added: PADS does not work ``straight out of the box''. The level of knowledge modelers are required to master is still too high. Some aspects such as causality constraints and data dissemination are hard to manage and understand.\\

Let's now investigate the simulation users more in detail. It is quite obvious that PADS techniques are, in some cases, necessary (and provide a benefit) sometimes they are not necessary at all (e.g.~when PADS is slower than sequential). Therefore, each time the main question should be: which is the better choice? There are many possibilities, as sequential/parallel/distributed only refer to the high level approaches. Yet we must remember that each of them can be implemented using different execution architectures (e.g.~multi-core CPUs, clusters, public or private clouds). Up to now, the whole problem is left to the simulation model developer (or the simulator user), who will not easily be a PADS expert. It feels like PADS tools are for initiates: a better approach would be to hide from the users all such technical details that should be on duty of software tools.\\

In Section~\ref{sec:pads} we have said that we have a parallel simulation when the execution nodes are connected by a low latency network (e.g.~a bus), and conversely a distributed one when the latency is higher (e.g.~a LAN, WAN or even Internet). This categorization is simple and clear but very inadequate: in the real world the execution architectures are much more complex and heterogeneous in terms of hardware, software and runtime conditions. Nowadays, if there is a dedicated cluster for simulations, then this is very likely made by some multi-core (and often multi-processor) hosts interconnected with some kind of network. More often, the simulations are run on spare servers or desktop PCs during nights and weekends. In Section~\ref{sec:challenges}, we predicted that, in the next years, the ``\emph{everything as a service}'' approach that is at the basis of cloud computing will hit PADS. If that is true, then the execution environment will turn out to be still more complex. For example, if cloud 
techniques are used to build private clouds, then it is possible to obtain some sort of control on such execution environment. In other words, in private environments it is much easier to guarantee performance and reliability than in services that are furnished by third-part service providers. Conversely, this is not conceivable in public clouds (such as Microsoft Azure, Amazon EC2 or Google App Engine). In these services, the user can only rely on some general Service Level Agreements (SLAs).\\

Nowadays, with current simulation technologies, the simulator user is left alone in choosing almost everything. Why is it so difficult to decide what is the best approach to be used? Even the ``sequential or PADS'' choice is hard to make because it depends on a very large set of dynamic parameters that can be found in all the logical layers of the architecture, starting from the simulation model behavior and down to the hardware performances. All those parameters need a case-by-case evaluation. Furthermore, they can also change within each simulation run, due to many different factors such as the semantic of the simulated model and the unexpected presence of background load in the execution architecture. It is unrealistic to think that the simulation user can tackle all of such aspects and details. It is the software that should be in charge of ``making it easy''.

	\subsection{Usability (lack of)}
%
% Espandere il tema dell'usabilità riprendendo quanto sopra ma in ottica più estesa
%
The user of a simulation tool should focus on the modeling aspects and on the analysis of the obtained results. In practice, it happens that the modeler uses a different tool if he wants to build a sequential simulation or a PADS one. Let's suppose that he chooses to build a sequential simulation and after completing the implementation of the model he discovers that the model is so complex that the simulator is too slow. Very likely, the transition to a PADS would result in the re-making of the most part of the simulator. For example, he would have to decide the partitioning of the simulation model.\\

It would be much better to maintain the implementation of the simulation model separate from all other aspects related to the implementation of the simulation (e.g.~synchronization, data distribution, partitioning, load balancing). In the past, this has been tried many times, but often with poor results. For example, it is very difficult to insulate the model from the synchronization management. For this reason, the Standard for Modeling and Simulation High Level Architecture (IEEE 1516)~\cite{ieee1516} supports optimistic synchronization, but the implementation of all the support mechanisms (such as roll-backs, see Section~\ref{sec:pads}) are left to the simulation modeler~\cite{Santoro2006}.\\

A more mature approach to modeling and simulation would require the software to manage all the low level details of the simulation in the more appropriate way, in order to obtain the better performance. It is worth noting that, as it will be clear in the following of this section, obtaining the simulation results as fast as possible is not always the best. Some other criteria have to be taken into account as well.

	\subsection{Cost assessments: the need for new metrics}
%
% Fare un'analisi generale dei costi legati al classico approccio PADS e notare che
% questi non vanno bene per una situazione di "everything as a service" dove spesso paghi
% in base all'utilizzo (es. il time-warp)
%
It is common to use the amount of time for completing a simulation run (Wall-Clock-Time, WCT) as the metric for evaluating the performance of a simulator. This is acceptable in a classic execution architecture, but it is not when computation and communication services are rented (e.g.~using the services offered by a public cloud provider). In this case, the cost follows the ``pay for what you use'' rule and the user of simulation tools will have to consider: 
\begin{itemize}
	\item how much time he can wait for the results;
	\item how much he wants to pay for running the simulation. 
\end{itemize}
It is clear that, in this scenario, every computation and communication overhead should be minimized or at least carefully scrutinized.\\

Are the current PADS mechanisms suitable for this new evaluation metric? The detailed analysis of the many different mechanisms and variants that are used for building PADS would require a very large amount of time and space, but we can focus on synchronization. As introduced in Section~\ref{sec:pads}, two main approaches have been proposed for implementing synchronization services in PADS: a) conservative and b) optimistic.\\

The Chandy-Misra-Bryant algorithm~\cite{misra86} is one of the most well-known ways for implementing conservative synchronization. Due to its nature, it needs to introduce in the simulation some artificial events (i.e.~without any semantic content) with the aim to make the simulation proceed and to avoid deadlock. The number of such events introduced by the synchronization algorithm can be very large~\cite{devries1990,rizvi2006}. In the years, many variants have been proposed to reduce the number of such events~\cite{Su:1988:VCD:865909}, but the amount of extra communications for their delivery can still be prohibitive.\\

The consideration that computation (i.e~CPUs) is much faster and cheap than communication (i.e.~communication networks) is at the basis of optimistic synchronization. This means that, in a distributed simulation, the CPUs will be very often idle, waiting for some data from the network (e.g.~the delivery of events). If such assumption is true, then it could be a good idea to compute events also if we are not sure that they are in the correct order. If they are not, then the simulation will roll-back to a correct state, wasting some computation and communication. It is clear that even this approach is not well suited for execution environments in which you ``pay for what you use''. In a optimistic simulation, a very large part of the computation can be thrown away due to roll-backs.\\

To summarize, the public cloud comes with a pricing scheme that is very different from the previous one, in which most of the budget was for the hardware. If the goal is to have simulations that really follow the new ``everything as a service'' paradigm, then we need some new mechanisms to implement PADS. Mechanisms that need to be less ``expensive'', both in terms of computation and communication requirements. In this case, the main evaluation metric is not the execution speed but the pricing scheme (or a combination of both). Furthermore, given that such price is always decided by the market, it could change very quickly.

	\subsection{In search of performance}
%
% Ribadire che non è solamente una questione di costo ma anche di performance della PADS,
% prendiamo ad esempio gli algoritmi di sincronizzazione:
%	conservativi
%		cmb: notoriamente il cmb, a causa dei null messages consuma molte risorse di comunicazione
%			questo non è positivo in uno scenario molto distribuito, un singolo nodo può diventare
%			facilmente un bottleneck in caso di rallentamento
%		timestepped: la velocità della simulazione è quella del suo componente più lento
%	ottimistico
%		timewarp: nato per evitare che i processori rimanessero senza fare nulla in attesa della
%			comunicazione. Notoriamente funziona bene quando i vari LP avanzano in modo
%			omogeneo. Questo può facilmente non avvenire in public cloud
%
% In un'architettura con una certa variabilità il rischio è quello di ottenere prestazioni miserabili
Let's go on with our discussion about the implementation of PADS in a public cloud execution environment and let's suppose that this time our main goal is to obtain the results as fast as possible. In other words, we will ignore the cost assessments introduced in the previous subsection.\\

As usual, for the sake of simplicity, we will focus on synchronization, even if it is obvious that many other aspects should be considered in detail. What happens if the synchronization algorithms described in Section~\ref{sec:pads} are run, without modifications, on a public cloud? What level of performance is it possible to expect?\\

We can start our analysis with the simplest synchronization algorithm: the timestepped. As said before, the simulation time is divided in a sequence of steps and it is possible to proceed to the next timestep only when all the components (i.e.~LPs) in the simulation have completed the current one. It is clear that the execution speed is bounded by the slowest component. This can be very dangerous in execution environments in which the performance variability is quite high. What about the Chandy-Misra-Bryant algorithm? We have already said that this algorithm is very demanding in terms of communication resources and that in this case a slow LP would become the bottleneck of the whole simulation. The last possibility is to use an optimistic synchronization algorithm such as the Jefferson's timewarp~\cite{timewarp}. This algorithm is not very promising either: timewarp is well-known to have very good performance when all LPs can proceed with an execution speed that is almost the same. This usually means that 
all LPs have to be very homogeneous in terms of hardware, network performance and load. Otherwise, the whole simulation would be slowed down by the roll-backs caused by the slow LPs. A requirement that is hard to satisfy in a public cloud environment.\\

As expected, the implementation of cloud-based PADS is not so simple as running the current tools on a public cloud. If performances and costs are important issues, then many parts of the current PADS approach have to be revised. In the following section, we will propose a new approach to deal with some of these issues.

\section{In the Search of Adaptivity: \\the ART\`IS/GAIA+ approach}
\label{sec:artisgaia}

Let's start with a warning: the ``silver bullet'' does not exist, even in simulation. The last attempt in PADS to obtain a ``one fits all'' solution has produced the IEEE 1516 - High Level Architecture (HLA) standard~\cite{ieee1516}, that is quite complex to use, lacks some basic features and has lead to many performance issues.\\

Our proposal for a different approach~\cite{gda-ijspm-2009} to the many problems described in this paper, involves first of all some work on the \emph{partitioning problem}. That is about decomposing the simulation model into a number of components and then properly allocating them among the execution units. This allocation procedure has at least two main goals to pursue: the computation load in the execution architecture has to be kept approximately balanced and in the meantime the communication overhead has to be minimized~\cite{bagrodia98}. If both these requirements are satisfied, then the execution is likely to be efficient. The hard part is that all of this has to be: i) \emph{transparent to users}, ii) \emph{dynamic and adaptive} (given that both the model behavior and the execution architecture conditions are not predictable). In other words, the runtime conditions are more and more often unpredictable and, moreover, the environment is dynamic and very heterogeneous. The direct consequence is that, 
in this case, all static (and analytical) approaches are not adequate.

\subsection{Model decomposition}
What we propose is the partitioning of the simulated model in very small parts (referred to as entities). Each entity represents a tiny piece of the simulated model and interacts with other entities to implement the model behavior. In this way, the execution architecture, that is composed of multiple nodes, is nothing more than a set of containers for the Simulated Entities (SE). In this case too, each container is called Logical Process, LP. In practice, the distributed simulation is organized as a Multi Agent System (MAS)~\cite{Wooldridge:2009:IMS:1695886}, a paradigm that has been demonstrated very easy to use, solid and promising. About our proposal, it is worth noting that the SEs are not statically allocated on a specific LP, but can be migrated with the aim to satisfy the partitioning constraints and to improve the runtime efficiency of the simulator~\cite{gda-dsrt-2004}.

\subsection{Dynamic partitioning}
More in detail, in managing the partitioning we have to consider two main aspects. Firstly, with respect to a sequential (i.e.~monolithic) simulation, every PADS has to deal with a significantly higher communication cost (e.g.~network latency and bandwidth limitations). Reducing this cost to the bare minimum is of main importance. Secondly, the simulator speed is bounded by its slowest component and therefore smart load balancing strategies should be implemented. To reduce the communication cost, the main strategy is to cluster the highly interacting SEs within the same LP~\cite{gda-pads-2003}. This clustering has the effect of increasing the use of low latency and high bandwidth networks (e.g.~the RAM within the host) and conversely reducing the usage of very costly communication technologies (e.g.~LAN, WAN, Internet). In practice, this can be obtained evaluating the communication pattern of each SE. It is pretty obvious that the other side of the problem is that clustering all the SEs in the same LP is (
usually) not a good load balancing strategy. Moreover, if a LP is overloaded, then it is going to slow-down the whole simulator and therefore the clustering is less important than load balancing. In other words, the partitioning is a very dynamic optimization problem with multiple goal functions and with a lot of parameters with unpredictable values.\\

As said before, the use of analytical methods to tackle such problem is unrealistic: we have to rely on heuristic methods. A set of heuristics is used to evaluate the simulator and the execution architecture step-by-step, and to decide if reallocations (of SEs) are necessary. Obviously migrations have a cost, due to the ``serialization'' of state variables (in SEs) and their network transfer. If it is favorable to cluster some SEs in the same LP, then some reallocations will be managed. That will also happen in case of imbalances in the execution architecture. All of this has to be done for the whole simulation length, given that both the execution architecture runtime conditions and the simulated model behavior will change with time. Some special cases should be analyzed more in detail. For example, if the amount of computation required by the simulation model is so low that no parallelization is necessary, then the execution architecture should automatically shrink up to a single LP (that is, a sequential 
simulation). Conversely, in case of a large amount of work the communication cost will be balanced by the benefit of parallel execution and therefore the number of LPs in the simulation needs to be increased.

\subsection{Finding and removing bottlenecks}
This approach is promising because it can help to solve many problems of PADS. For example, using an instrumented version of the synchronization algorithm (e.g.~a modified form of timestepped synchronization) it is possible to detect which LPs are slow in the simulation execution. As said before, the execution speed of PADS is usually bounded by its slowest component. Finding the bottlenecks is the first part of the solution, while the second part is once again based on the migration of SEs. If a LP is slow with respect to the other parts of the simulations, that means that: a) it is overloaded or b) the communication network of the LP is introducing too much delay. In both cases there can be many different reasons, for example the LP overload can be due to the semantic of the simulation model, to the presence of some background load in the host that is running the simulation, or even to the hardware characteristics of the host. In all such cases, the solution is to unload part of the SEs from the slow LP. 
In practice, this can be done migrating some SEs from the slow LP to a faster one. In this way, the bottlenecks can be removed, but it is necessary to continue this reallocation strategy for the whole simulation length. In many cases, the final result can be a ``smooth'', faster and more effective execution~\cite{gda-ijspm-2009}.

\subsection{ART\`IS and GAIA+}
In the last years, we pursued the approach described in the previous section with the implementation of a new simulation middleware (called Advanced RTI System, ART\`IS) and the companion GAIA+ framework (Generic Adaptive Interaction Architecture)~\cite{gda-ijspm-2009,gda-dsrt-2004,gda-pads-2003}.\\

While this effort is still not complete, a lot of different systems and scenarios have been evaluated to validate our approach. For example, both wired and wireless communication environments have been studied~\cite{gda-simutools-09,moves}. Using some of the features previously described it has been possible to manage the fine grained simulation of complex communication protocols (e.g.~IEEE 802.11) when in presence of a huge number of nodes (up to 1 million)~\cite{gda-ijspm-2009}. In the wired case, we worked on the design and evaluation of gossip protocols in unstructured networks (e.g.~scale-free, small-world, random)~\cite{gda-disio-10,gda-disio-11}. In all of these cases, we obtained a very good performance increase with respect to traditional simulation techniques. The adaptive techniques implemented in the simulation middleware have permitted the usage of low cost commercial off-the-shelf hardware. \\

Testing our prototype implementation, we learned that adaptation is a key component for performance and that all the proposed adaptive mechanisms can be implemented using very simple heuristics. The results demonstrate that very often simple approaches are much more effective than complex ones.\\

The ART\`IS middleware, the GAIA+ framework, a set of simulation models, the support scripts and the scenario definition files used in many research papers that we published in the last years, are freely available, for research purposes, on the research group website~\cite{pads}. A large part of the software is provided in both binary and source code versions. We expect to release all the source code in the next years, using an open source license.

\subsection{ReliableGAIA+}
As said in the previous subsection the work on GAIA+ and ART\`IS is still in progress. One of our main research lines is about fault-tolerance in distributed simulation. Usually, if a LP crashes (e.g.~for hardware issues or network problems) the result is the failure of the whole simulation run. We aim to build PADS that are able to continue the simulation at the cost of some computation and communication overhead. For example, with the aim of running PADS on top of unreliable cloud-based services (as described in Subsection~\ref{subsec:service}), following our approach based on the migration of SEs.\\

To provide fault-tolerance, each SE is implemented as a set of replicated Virtual SEs (VSEs) and also in this case the VSEs will be migrated in the execution architecture to improve load balancing and communication. Given that each VSE will be executed in a different LP, the crash of a part of the execution architecture will be no more a fatal problem. Tuning the number of replications of each SE can provide a different level of fault-tolerance. We are currently investigating the low level details of this proposal and implementing the necessary modifications in the simulation middleware. In case of success, the performance evaluation of ReliableGAIA+ will start in the next months.

\section{Conclusions}
\label{sec:conc}

In this paper we have provided a short introduction to the Parallel and Distributed Simulation (PADS) main concepts and we have seen that current PADS technologies are unable to fulfill many requirements (e.g.~usability, adaptivity). Furthermore, new challenges such as the ``many cores architecture'' and the ``simulation as a service'' are emerging. In the next years, most computers will be equipped with many cores CPUs and the cloud-based technologies will be commonly used. In particular, the public cloud will be an execution environment in which the computation and communication resources can be obtained on demand and paid for only for the capacity that is actually used. Our analysis shows that current PADS techniques are unable to fit well with these new execution architectures and that a lot of work is needed to increase usability and performance of simulators in such conditions.\\

We claim that a solution for such problems has to deal with the partitioning of the simulation model among the execution nodes. With this aim, we proposed an approach based on multi-agent system. Its main characteristic is the adaptive migration of the simulated entities between the execution units. Mechanisms based on heuristics can be implemented for both reducing the communication cost and load balancing the simulation. Our proposal has been implemented in the ART\`IS/GAIA+ simulation middleware and tested with many models, and the obtained performance outcomes are very promising.

\section*{Acknowledgments}
The author wish to acknowledge the helpful comments and criticisms of Lorenzo Donatiello, Kurt Vanmechelen and Martina Brini to the view described in this paper. Needless to say, the responsibility for the content and the views expressed in this paper remains with the author.

%%%%%%%%%%%%%%%%%%%%
%% BIBLIOGRAFIA
%%%%%%%%%%%%%%%%%%%%

\small{
\bibliographystyle{abbrv}
\bibliography{paper}  
}

% \balancecolumns

\end{document}